\documentstyle[11pt,paspconf,epsf]{article}
\begin{document}

\title{The David Dunlap Observatory IVC Distance Project: First Results} 

\author{T.~E. Clarke, Gabriela Mall\'en-Ornelas, Marcin Sawicki, Mark
  Brodwin, Rosemary McNaughton, Michael D. Gladders, Christopher R.
  Burns and A. Attard}

\affil{Department of Astronomy, 60 St.\ George Street, University of
Toronto, M5S 3H8, Canada}

\begin{abstract}

  We present distance estimates to a set of high-latitude
  intermediate-velocity HI clouds. We explore some of the
  physical parameters that can be determined from these results, such
  as cloud mass, infall velocity and height above the Galactic plane.
  We also briefly describe some astrophysical applications of these
  data and explore future work.

\end{abstract}


\section{Intermediate Velocity Clouds}

Clouds of neutral hydrogen at high galactic latitudes are detected as
discrete velocity structures in 21cm line emission. The population of
intermediate velocity clouds (IVCs) has $|V_{LSR}|$ $<$ 70 ${\rm km \ 
  s^{-1}}$. The nature of the clouds remains unknown, although several
scenarios for their origin have been proposed (see Wakker \& van
Woerden [1997] for a review). In order to understand the origins and
dynamics of these clouds, it is necessary to have accurate distance
estimates to them.

We have undertaken a project to measure the distances to high-latitude
intermediate velocity clouds selected from the Heiles, Reach \& Koo
(1988) sample. We observe a set of target stars along the lines of
sight to the clouds and use interstellar absorption features to
determine if the stars are behind or in front of the clouds. From
spectral classification, we deduce the distances to these foreground
and background stars and hence bracket the distances to the clouds. In
this paper, we explore some potential applications of our data.

\section{Some Applications of Distance Measurements}

\begin{figure}
\plotone{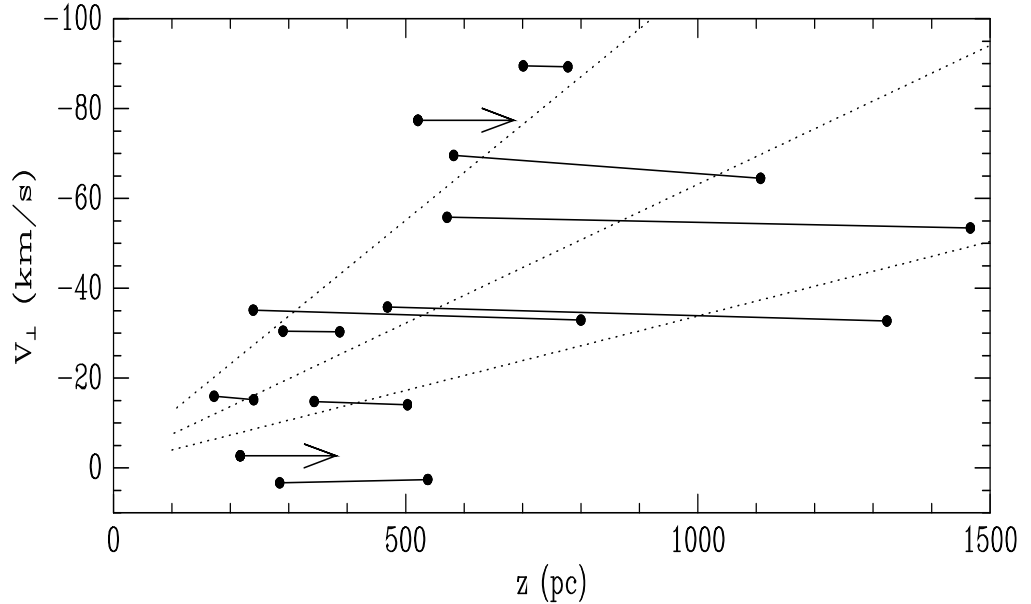}
\caption{Velocity perpendicular to the Galactic plane plotted as a function 
  of cloud height above the plane. The connected symbols indicate the
  upper and lower brackets computed from the line of sight distance
  estimates. Perpendicular velocities were calculated assuming that
  the clouds' motions are directed vertically (in the co-rotating
  Galactic frame), and include corrections for the differential
  Galactic rotation. Note that, in general, the higher clouds are
  falling faster.  The dotted lines correspond to the predictions of
  the terminal velocity models of Benjamin \& Danly (1997); top to
  bottom lines correspond to cloud densities of $10^{20}$,
  $3\times 10^{19}$ and $10^{19}$ ${\rm cm^{-2}}$ in their warm HII +
  HI + hot halo model. Our data appear to be consistent with these
  terminal velocity infall models.\label{fig-1}}
\end{figure}

We have upper and lower distance estimates for 9 clouds and lower
limits only for another 2 clouds (Gladders {\it et al.}, 1998a,b; Burns
{\it et al.}, 1999; see Table~\ref{tbl-1}).  These distances allow us
to calculate a number of physical parameters of the clouds. For
instance, we can calculate the infall velocity ($v_\perp$) and cloud
distance above the Galactic plane (z) as shown in Figure~\ref{fig-1}.
These perpendicular velocities can be compared to theoretical
predictions of ballistic infall models and terminal velocity models,
and appear to be consistent with the latter.

Distance estimates also enable us to calculate the cloud mass.
Table~\ref{tbl-1} shows the cloud HI mass estimates obtained
by combining our distance brackets with the HI column
densities from Heiles, Reach \& Koo (1988) and angular sizes from IRAS
100 $\mu$m dust images.

\begin{table}
\caption{IVC distance brackets and HI mass estimates.} \label{tbl-1}
\begin{center}\scriptsize
\begin{tabular}{||l|l|r@{--}l|r@{--}l||} \tableline
\multicolumn{6}{||c||}{} \vspace{-0.3cm}\\ 
\parbox{2in}{\center{{\bf CLOUD(S)}}}&\parbox{1in}{\center{{\bf
V$_{\mbox{{\tiny LSR}}}$\tablenotemark{1}}\ \  (km/sec)}}&\multicolumn{2}{|c|}{\parbox{0.5in}{\center{{\bf d } (pc)}}}&\multicolumn{2}{|c||}{\parbox{0.8in}{\center{{\bf M$_{\mbox{{\tiny HI}}}$} (M$_\odot$)}}}\\
&&\multicolumn{2}{|c|}{}&\multicolumn{2}{|c||}{} \vspace{-0.05cm}\\ \tableline %\tableline
&&\multicolumn{2}{|c|}{}&\multicolumn{2}{|c||}{} \vspace{-0.2cm}\\
G163.9+59.7 &$-19.0$&300&2380& ~~~~~~~5&250\\ 
G139.6+47.6 \& G141.1+48.0&$-12.1$ \& $-12.9$& 120&420&10&110\\
G135.5+51.3 &$-47.2$&310&1900&100&3560\\
G149.9+67.4 &$-6.3$&260&660&5&20\\
G249.0+73.7 &$-0.6$&\multicolumn{2}{|r|}{$\geq \!\!180$~~~}&\multicolumn{2}{|c||}{~$\geq\!\! 60$}\\
G124.1+71.6 &$-11.4$&240&2020&50&3430\\
G107.4+70.9 \& G99.3+68.0 & $-29.9$ \&  $-26.6$&530&1220&220&1170\\
G86.5+59.6&$-39.0$&\multicolumn{2}{|r|}{$\geq\!\! 430$~~~}&\multicolumn{2}{|c||}{~~$\geq \!\!380$}\\
G90.0+38.8 \&  G94.8+37.6 (Draco)\tablenotemark{2}&$-23.9$ \& $-23.3$& 330&860&190&1340\\
G81.2+39.2 &$+3.5$&320&1260&270&4180\\
G86.0+38.3&$-43.4$&640&3560&130&4070\\ \tableline 
\end{tabular}
\end{center}
\tablenotetext{1}{Heiles, Reach and Koo (1988)}
\tablenotetext{2}{Presented in detail in Gladders {\it et al.}, 1998a}
\end{table}

\section{Future Work}

We are currently extending our project to include more intermediate
velocity clouds. We are also refining (where possible) the above cloud
distances by observing more stars and by obtaining spectroscopic
classification for those stars which currently have only colour-based
distances.

Future high resolution observations of several strong interstellar
absorption line features in background stars will allow us to
constrain cloud metallicities.  The metallicity will be relevant to
the question of the origin of the clouds (i.e., whether they are
Galactic ejecta or primordial gas).

IVC distances can also be used to constrain the nature of the soft
X-ray background radiation (Kerp, 1996).  The origin of this emission
is considered to be from either the local interstellar medium or from
the galactic halo. The detection of an ``X-ray shadow'' toward the
Draco cloud (Snowden {\it et al.}, 1990) provided evidence that at
least some of the emission is from large distances. Further soft X-ray
shadowing measurements toward more clouds with distance brackets will
provide insight into the origin of this emission.

\acknowledgments

T.E.C.\ would like to thank the local organizing committee for
financial assistance to attend the workshop. We wish to acknowledge
Wayne Barkhouse, L.\ Felipe Barrientos, M.P.\ Casey, Jason Clark, Devon
Hamilton, J.L.\ Karr, Stefan W.\ Mochnacki, Sara M.\ Poirier and Marcelo
Ruetalo-Pacheco, who gathered part of the observations for this
project.  We thank the David Dunlap Observatory (DDO) for very generous
observing time allocation.

\end{document}